\documentclass[11pt]{article}

\usepackage{verbatim}
\usepackage{amsmath}
\usepackage{amsfonts}
\usepackage{amssymb}
\usepackage{bbm}
\usepackage{latexsym}
\usepackage{graphics}
\usepackage{lscape} 
\usepackage{epsfig}
\usepackage{color}

\usepackage{cite}

\renewcommand{\d}{\mathrm{d}}

\newcommand{\captn}[1]{\vspace{-3ex}\caption{\small #1}}

\DeclareMathSymbol{\mg}{\mathrel}{symbols}{"1D}

\newcommand{\gvf}{\varphi}

\newcommand{\gm}{\mu}
\newcommand{\gn}{\nu}

\newcommand{\go}{\omega}

\newcommand{\gp}{\pi}

\newcommand{\gF}{\Phi}

\newcommand{\gO}{\Omega}

\newcommand{\gPs}{\Psi}

\newcommand{\cF}{{\cal F}}

\newcommand{\cM}{{\cal M}}

\newcommand{\ui}{{\underline i}}

\newcommand{\tr}{\text{tr}}

\newcommand{\ra}{\rightarrow}

\newcommand{\Kh}{K\"{a}hler}
\newcommand{\beq}{\begin{equation}}
\newcommand{\eeq}{\end{equation}}
\newcommand{\barr}{\begin{array}}
\newcommand{\earr}{\end{array}}
\newcommand{\equ}[1]{\begin{gather} #1 \end{gather}}

\newcommand{\enums}[1]{\begin{enumerate} #1 \end{enumerate}}

\newcommand{\arry}[2]{\begin{array}{#1} #2 \end{array}}

\newcommand{\non}{\nonumber}
\newcounter{oldcounter}

\newcommand{\Intr}{\mathbb{Z}}
\newcommand{\Cplx}{\mathbb{C}}

\newcommand{\ba}[2]{\[\begin{array}{#2}\label{#1}}
\newcommand{\ea}{\end{array}\]}
\newcommand{\be}{\begin{equation}}
\newcommand{\ee}{\end{equation}}
\newcommand{\bea}{\begin{eqnarray}}
\newcommand{\eea}{\end{eqnarray}}

\newcommand{\U}[1]{\mathrm{U(#1)}}

\newcommand{\SO}[1]{\mathrm{SO(#1)}}

\newcommand{\rep}[1]{\mathbf{#1}}
\newcommand{\crep}[1]{\overline{\rep{#1}}}

\newcommand{\sm}{{\,\mbox{-}}}

\def\ttle{Multiple anomalous U(1)s in heterotic blow--ups}

\def\abstrct{
The existence of multiple anomalous U(1)s is demonstrated explicitly
in a blow--up version of a heterotic $\Intr_3$ orbifold. Another
blow--up of the same orbifold supports further evidence for the
type--I/heterotic duality in  four dimensions. It has a single
anomalous U(1) which does not factorize universally. As multiple
anomalous U(1)s as well as non-universal factorization have never been
established on heterotic orbifolds explicitely, these findings might
appear contradictory at first sight. Possible inconsistencies are
avoided by reinterpreting a charged twisted state as a second
non--universal localized axion. The mismatch between the charges of
the orbifold and blow--up spectra is resolved by suitable field
redefinitions. The anomaly of the field redefinitions corresponds to
the difference of blow--up and heterotic orbifold  anomalies.   
}

\begin{document}

\thispagestyle{empty}

\begin{flushright}
HD-THEP-07-07 \\
SIAS-CMTP-07-2 \\ 
hep-th/yymmnnn\\
\end{flushright}
\vskip 2 cm
\begin{center}
{\Large {\bf \ttle} 
}
\\[0pt]

\bigskip
\bigskip {\large
{\bf S.\ Groot Nibbelink$^{a,b,}$\footnote{
{{ {\ {\ {\ E-mail: grootnib@thphys.uni-heidelberg.de}}}}}}},
{\bf H.P.\ Nilles$^{c}$\footnote{
{{ {\ {\ {\ E-mail: nilles@th.physik.uni-bonn.de}}}}}}}
{\bf M.\ Trapletti$^{a,}$\footnote{
{{ {\ {\ {\ E-mail: M.Trapletti@thphys.uni-heidelberg.de}}}}}}},
\bigskip }\\[0pt]
\vspace{0.23cm}
${}^a$ {\it 
Institut f\"ur Theoretische Physik, Universit\"at Heidelberg, \\ 
Philosophenweg 16 und 19,  D-69120 Heidelberg, Germany 
\\} 
\vspace{0.23cm}
${}^b$ {\it 
Shanghai Institute for Advanced Study, 
University of Science and Technology of China,\\ 
99 Xiupu Rd, Pudong, Shanghai 201315, P.R.\ China
 \\} 
\vspace{0.23cm}
${}^c$ {\it 
Physikalisches Institut der Universit\"at Bonn, \\
Nussallee 12, 53115 Bonn, Germany
 \\} 
\bigskip
\end{center}

\subsection*{\centering Abstract}

\abstrct

\newpage 
\setcounter{page}{1}

\def\brkline{}
\def\brkend{\\[1ex]}


\def\Models{
\begin{table}
\begin{center} 
\scalebox{1}{
\begin{tabular}{| l | c | l |}
\hline && \\ [-2ex]
Model & Gauge Group & \multicolumn{1}{c|}{Spectrum }
\\[1ex]\hline\hline && \\ [-2ex]
Het.O. & 
$\SO{8} \!\times\! \U{12}$ & 
$\arry{c}{
\frac 19\, (\rep{8},\rep{12})_1 
+ \frac 19\, (\rep{1}\,, \crep{66})_{\sm 2} 
\brkline
+ (\rep{1},\rep{1})_4 + (\rep{8}_+, \rep{1})_{\sm 2}
}$
\brkend \hline & & \\ [-2ex]
TypeI & 
$\SO{8} \!\times\! \U{12}$ & 
$\arry{c}{\frac 19 (\rep{8},\rep{12})_1 
+ \frac 19 (\rep{1}, \crep{66})_{\sm 2}}$
\\[1ex]\hline & & \\ [-2ex]
U(1)$^2$ & 
$\U{4} \!\times\! \U{12}$ & 
$\arry{c}{
\frac 19 (\rep{4},\rep{12})_{1,1} +
\frac 19 (\crep{4},\rep{12})_{1,\sm 1 } 
\brkline
+\frac 19 (\rep{1},\crep{66})_{\sm 2, 0}
+(\rep{6},\rep{1})_{0,2}
}$ 
\brkend\hline
\end{tabular}}
\end{center} 
\captn{The spectra of the heterotic $\Intr_3$ orbifold
and two blow--ups are displayed: The spectrum of ``TypeI''
equals the type--I $\Intr_3$ orbifold, and  
``U(1)$^2$'' has  two anomalous U(1)s.}
\label{tb:Models}
\end{table}}

\section{Introduction and results}

For phenomenological applications of heterotic string
compactifications it is important to know the number of anomalous
U(1)s~\cite{Dine:1987xk,Atick:1987gy}. It has been generally 
accepted that heterotic orbifold
models~\cite{dixon_85,Dixon:1986jc,ibanez_87,ibanez_88} 
contain at most a single anomalous U(1). Recent 
publications~\cite{Blumenhagen:2005pm,Blumenhagen:2005ga} argued that
the low energy limit of the heterotic string, i.e.\ super
Yang--Mills (YM) theory coupled to supergravity, on general smooth
Calabi--Yau(CY)s with U(1) bundle backgrounds multiple anomalous
U(1)s are possible. Assuming that such CYs have continuous singular
orbifold limits, the existence of multiple anomalous U(1)s seems
puzzling. In 6D matching between heterotic orbifolds and
blow--ups seems always possible~\cite{Honecker:2006qz}. To investigate
this issue in 4D, explicit blow--ups of the orbifold fixed  points
would be very instructive. Since for the $\Cplx^3/\Intr_3$ orbifold
the blow--up and its U(1) bundles have recently been
constructed~\cite{Nibbelink:2007rd} (see also~\cite{Ganor:2002ae}), we
would like to focus on anomalous U(1)s in such blow--ups.

The following obstructions seem to prevent a straightforward
identification of heterotic orbifold models and their blow--up
counterparts:  
\enums{
\item In contrast to heterotic orbifolds, blow--ups can have more than
one anomalous U(1).  
\item While orbifold theories seem to contain at most a single
axion relevant for Green-Schwarz(GS) anomaly cancellation,
blow--up models can have multiple axions. 
\item In blow--up models with a single anomalous U(1) the anomalies
do not factorize universally as they do in heterotic orbifold models. 
\item The spectra of orbifolds and their corresponding blow--ups 
often do not match.
\item Even when the non--Abelian spectra agree, 
the U(1) charges of twisted states seem never to be identical. 
} 
We argue that all these discrepancies can be explained by suitable field
redefinitions. In particular, we reinterpret a charged twisted singlet
on the orbifold, that drives the blow--up by taking Vacuum Expectation
Value(VEV), as a localized axion. It takes part in a non--universal
GS anomaly cancellation, very much like twisted RR--axions in
type--I models. The anomaly in these field redefinitions precisely
corresponds to the difference between blow--up and heterotic orbifold
anomalies.

We first describe this procedure for $\Cplx^3/\Intr_3$ blow--up models
with U(1) bundles in general. After that we inspect two concrete
blow--up models of a specific heterotic $\Intr_3$ orbifold to illustrate
some details. The orbifold model has been considered
before~\cite{Kakushadze:1997wx,Kakushadze:1998cd} as evidence for
heterotic/type--I duality~\cite{Polchinski:1995df} in 4D. The 
second example constitutes a different blow--up of the same orbifold
that has two anomalous U(1)s.

\section{Multiple Anomalous U(1)s in Blow--Up}

In~\cite{Nibbelink:2007rd} the blow--up $\cM^3$ of the $\Cplx^3/\Intr_3$
orbifold and its line bundles are described. Using these results we
decompose the 10D GS 2--form $B_2$ into 4D perturbations as   
\equ{
B_2 ~=~ b_2 ~+~ i\cF_2\, b_0 ~+~ \go_2\, B_0~. 
\label{expandB2}
}
Both the U(1) bundle field strength $i \cF_2$  and the \Kh\ form
$\go_2$ are harmonic 2--forms. (For the \Kh\ form this follows
automatically, while for the U(1) field strength this is guaranteed by
the Hermitean YM  equations.) Aside from the 4D anti--symmetric tensor
$b_2$, dual to the universal axion $a^{uni}$, the 2--form $B_2$ is
decomposed into two 4D scalars $b_0$ and $B_0$. The $B_0$ is
interpreted as the 6D internal part of $B_2$, i.e.\ the nine untwisted
$B_{\ui j}$, $\ui, j = 1,2,3$, because $\go_2$ reduces to the orbifold
\Kh\ form in the blow down limit. The state $b_0$ is localized near
the orbifold singularity, since $i\cF$ becomes strongly peaked there
in that limit~\cite{Nibbelink:2007rd}. The expansion~\eqref{expandB2}
is complete:  As each of the nine  untwisted $B_{\ui j}$ states
contributes $1/27$ at a $\Intr_3$ fixed point~\cite{Gmeiner:2002es},
the blow--up has $h_{11} = 9/27 + 1 = 4/3$, which indeed equals half its
Euler number~\cite{Nibbelink:2007rd}.

Which of the $b_2$, $b_0$ and $B_0$ mediate GS mechanisms is
determined by their anomalous variations, that leave  the 3--form
field strength $H_3$ of $B_2$ invariant. Expanding $H_3$ in 
these 4D perturbations gives 
\equ{
H_3 = \d b_2 - \gO^{YM}_3  + \gO^L_3
 + i\cF_2 ( \d b_0 \!-\!  iA_V) 
+ \go_2 \ \d B_0. 
}
The usual anomalous variations of $b_2$ are found by restricting the YM and
gravitational Chern--Simons (CS) 3--forms, $\gO^{YM}_3$ and $\gO^L_3$,
to 4D. Since the expansion of 10D YM CS also contains a linear term in 
$i\cF \, H_V$ with $H_V = V^I H_I\,$,  the anomalous variation of
$b_0$ is determined by  $i A_V = \tr[H_V iA_1]$. The $H_I$ are the
Cartan generators of 10D gauge group and $V^I$ are numbers
characterizing the U(1) bundle~\cite{Nibbelink:2007rd}. Contrary $B_0$
never has an anomalous variation, because neither CS 3--forms have
decompositions proportional to $\go_2$.

The 4D anomaly $I_6$ is obtained from the factorized 10D anomaly
polynomial $I_{12} = X_4\, X_8$, where  
$X_{4} = \tr(iF_2)^2 - \tr\, R^2$ and $X_8$ is given
in~\cite{Green:1984sg,Green:1985bx}. By integrating over the blow--up we
get 
\equ{
I_6 ~=~ \int_{\cM^3} I_{12} 
~=~ \int_{\cM^3}  X_{2,2}\ X_{4,4}~+~ X_{0,4}\ X_{6,2}~, 
\label{SumFact}
}
where the first (second) index specifies the number of 6D internal
(4D Minkowski) indices of the forms. A third possible contribution,  
$X_{4,0}\ X_{2,6}$, integrates to zero due to the background Bianchi
identity. The integral of the first term can give rise to more than
one term in general. However, for the U(1) bundles on the blow--up of
$\Cplx^3/\Intr_3$ we get  
\equ{
X_{2,2} = 2\, i\cF_2 \, \tr (H_V iF_2)~, 
}
which ensure that this gives a single factor. Factorization in
heterotic orbifolds only occurs through the last term in~\eqref{SumFact}. 
Hence the blow--up of $\Cplx^3/\Intr_3$ can support at most two
anomalous U(1)s, and we can interpret $b_0 = a^{non}$ as a
non--universal axion. Together with the universal axion 
$a^{uni}$ (4D dual to $b_2$) it will take part in the 
GS mechanism of anomaly cancellation.

\section{Spectral matching}

Above we argued that on the blow--up of $\Cplx^3/\Intr_3$ orbifold there
can be two anomalous U(1)s. In this section we explain why there is no
contradiction with the general finding that heterotic orbifolds 
have at most a single anomalous U(1). Consider a heterotic orbifold
model and a proposal for a corresponding blow--up. Because the heterotic
orbifold has an anomalous U(1), the vacuum is unstable and some
twisted fields need to get non--vanishing VEVs. Depending on which
fields attain VEVs this is accompanied by further symmetry breaking to
the blow--up gauge group. Some twisted fields decouple by VEV induced
superpotential mass terms; this makes the non--Abelian spectrum of the
heterotic orbifold equal to that of the blow--up
model~\cite{Kakushadze:1997wx,Kakushadze:1998cd}. But this does not
resolve why charges of the fields present on both sides can be
different.

A possible mismatch of U(1) charges can be understood by analyzing the
consequences of fields with non--vanishing VEVs. Let $\gPs_q$ be a
twisted singlet chiral superfield, w.r.t.\ the unbroken blow--up gauge group
with U(1) charges $q = (q_1, \ldots, q_n)$, and $\gF_Q$ another
twisted chiral superfield, not necessarily a singlet, with charges
$Q$. A non--vanishing VEV $v$ of $\gPs_q$ means that in the quantum
theory this field can be represented as an exponential, and we can
redefine $\gF_Q$ with an arbitrary power $r$ of this exponential: 
\equ{
\gPs_q ~=~ e^{T}\, v~,
\quad 
\gF_Q ~=~ e^{r\,T} \, \gF_{Q'}~. 
\label{redgPsgF}
}
The new superfields $T$ and $\gF_{Q'}'$ transform under the U(1)
gauge transformations as: 
\equ{
T ~\ra~ T ~+~ iq_i \gvf_i~,  
\quad 
 \gF_{Q'}' ~\ra~ e^{ i (Q_i - r q_i) \gvf_i}\, \gF_{Q'}'
~, 
\label{gaugeT} 
}
where $\gvf_i$ are the gauge parameters of the U(1)s. Hence the
imaginary part $a_T$ of $T$ is related to the non--universal 
axion $a^{non}$ of the blow--up theory. In addition, we have obtained a
superfield $\gF_{Q'}'$ with U(1) charges $Q' = Q-rq\,$. We claim that 
it is always possible to perform such field redefinitions to make the 
charges of the non--decoupled twisted states equal to their blow--up
counterparts. Moreover, the fields that decouple can be given gauge
invariant superpotential mass terms using field redefinitions.

Next we explain how these field redefinitions help to get agreement 
between the anomalies of the heterotic orbifold and the blow--up models.
Upto this point $a_T$ only has anomalous gauge
variations~\eqref{gaugeT} but no anomalous couplings; still at this
stage only one axion  is involved in the GS mechanism. We call this
the heterotic axion $a^{het}$. Its anomalous couplings are determined
by the anomaly polynomial,  that factorizes universally 
$I_6^{het} = X_{2}^{het} \, X_{4}$. 
Here the GS 4--form $X_{4}$ is restricted to 4D, and
$X_{2}^{het}$, via the descent equations, determines the anomalous
variation of $a^{het}$. The anomalous couplings for the axion $a_T$
arises because the path integral measure is not invariant under the
anomalous field redefinitions~\eqref{redgPsgF}. The resulting
couplings can be deduced from the anomaly polynomial $I_6^{red}$
associated with the field redefinitions. It factorizes as 
$I_6^{red} = q_i i F^i_2 \, X^{red}_{4},$ 
where $iF_2^i$ denote the U(1) field strengths, because all field
redefinitions~\eqref{redgPsgF} involve only $T\,$. 
This anomaly combined with the heterotic anomaly 
$I_6^{het}$ equals the anomaly of the blow--up model: 
\equ{
I_6^{het} ~+~ I_6^{red} ~=~ I_6^{blo} ~=~
I_6^{uni} ~+~ I_6^{non}~.
\label{AnomSum}
} 
This equation reflects that after the field redefinitions the chiral
spectra of the heterotic orbifold and the blow--up models become
identical.

Finally we would like to understand the precise relation between the
heterotic axion $a^{het}$ and the axion $a_T$ obtained from the
superfield $T$ used in the rescaling~\eqref{redgPsgF}, and the two
blow--up axions, $a^{uni}$ (the  4D dual of $b_{\gm\gn}$) and $a^{non} =
b_0$. The anomaly polynomials $I_6^{uni}$ and $I_6^{non}$ of the
latter two are determined by the anomalous couplings of the zero modes
of $B_2$ 
\equ{
\int B_2\, X_8 ~=~ \int b_2\, X_{2,6} 
~+~ (b_0 \,i\cF_2 + B_0\,\go_2) X_{4,4}~. 
} 
They factorize as 
\equ{
(2\gp)^2\, I_6^{uni} = X_{2}^{uni}\, X_{4}~, \quad  
(2\gp)^2\, I_6^{non} = X_{2}^{non}\, X_4^{non},
} 
where $X_{2}^{non} = -\frac 1{96} \tr\, [ H_V iF_2]$ and 
\equ{ 
X^{uni}_{2} \,= \int_{\cM^3} \! \frac{X_{2,6}}{96 (2\gp i)^3}~,~
X_4^{non} \,=\, -\!\!\int_{\cM^3}\!\!\! \frac{2i\cF_2\, X_{4,4}}{(2\gp i)^3}~. 
}
Hence each term in~\eqref{AnomSum} can be computed independently,
providing a consistency check. The relation between the axions is now
fixed by noting that the anomalous variations of the axions $a^{het}$, 
$a_T$, $a^{uni}$ and $a^{non}$ are determined from $X_{2}^{het}$,
$q_i \, iF_2^i$, $X_{2}^{uni}$ and $X_{2}^{non}$, respectively, via 
the standard descent equations. Therefore, the sum relation for the 
anomalies~\eqref{AnomSum} implies that the interactions of the axions
with gauge and gravitational fields are related via
\equ{
a^{het}\, X_{4} ~+~ a_T \, X^{red}_4 ~=~ 
a^{uni} \, X_{4} ~+~ a^{non}\, X_4^{non}~. 
}
As the unbroken gauge group typically consists of  two or more group 
factors, this leads to an over constrained system of linear equations
relating $(a^{uni},a^{non})$ to $(a^{het}, a_T)$. Because both
$a^{het}$ and $a^{non}$ multiply $X_4$, we find 
\equ{
a^{uni} ~=~ a^{het} ~+~ c \, a_T~,    
\quad 
a^{non} ~=~ d \, a_T~, 
}
where $c$ and $d$ are model dependent constants.

To summarize we have shown that the chiral spectra of the heterotic
orbifold and blow--up models are identical upon using field
redefinitions that allow one to modify U(1) charges. In particular, a
charged twisted singlet on the heterotic orbifold is reinterpreted as
a localized axion in the blow--up theory. The difference between their
anomalies is precisely canceled by the anomalous variation of
this localized axion.

\section{A heterotic $\boldsymbol{\SO{32}}$ $\boldsymbol{\Intr_3}$
  orbifold}

\Models

We illustrate our general findings by considering two U(1) bundle
blow--ups of the heterotic $\SO{32}$ $\Cplx^3/\Intr_3$ orbifold model
with gauge shift $\frac 13(0^4,1^8,\sm 2^4)$. The gauge group and the
spectrum of this heterotic model~\cite{Giedt:2003an,Choi:2004wn,Nilles:2006np}
are given in the first row of table~\ref{tb:Models}. Since a
multiplicity factor of $\frac 19$ signals untwisted
modes~\cite{Gmeiner:2002es}, the  
model contains two twisted states: a singlet  $(\rep{1},\rep{1})_4$
and a $\SO{8}$ spinor $(\rep{8}_+, \rep{1})_{\sm 2}$ 
with charges $4$ and $\sm 2$, respectively. As usual for
heterotic models the anomaly polynomial factorizes universally  
\equ{
(2\gp)^2 I_6^{het} ~=~ - \frac 13 iF_1\, X_4~, 
\\[1ex] \non 
X_4 ~=~ 24\, (iF_1)^2 + 2\, \tr(iF_{12})^2 
+ \tr(iF_8)^2 - \tr\,R^2~,
}
where $F_1\,$, $F_{12}\,$ and $F_{8}$ denote the gauge field strengths
of $\U{1}\,$, $\U{12}$ and $\SO{8}\,$, respectively. 
The blow--ups are determined by VEVs of the twisted states that
satisfy the D-- and F--flatness  conditions. The D--flatness implies
that at least the singlet has a VEV to cancel the one--loop
Fayet--Iliopolous term~\cite{Dine:1987xk,Atick:1987gy}. (In addition,
if the spinor takes a VEV, two of its components are non--vanishing.)
Assuming that none of the untwisted states have VEVs, the relevant
part of the superpotential reads
\equ{
W ~\sim~ (\rep{1},\rep{1})_4 [(\rep{8}_+, \rep{1})_{\sm 2}]^2~. 
%
\label{superP} 
}
%
When both $(\rep{1},\rep{1})_4$ and $(\rep{8}_+, \rep{1})_{\sm 2}$
have VEVs, F--flatness is assured through the
presence of higher superpotential terms. The superpotential
(being a complex function of these fields) allows solutions 
with vanishing F-terms (and D-terms) at isolated points in
parameter space.

This heterotic orbifold model has received quite some attention in the
past because this model was suggested to have a type--I  $\Intr_3$
orbifold model
dual~\cite{Angelantonj:1996uy,aldazabal_98,Kakushadze:1997wx,Kakushadze:1998cd}.
However, because the GS anomaly cancellation in both models is
mediated by different fields, it had been questioned whether these
models can really be dual to each other~\cite{Lalak:1999bk}. 
Applying the general
formalism we developed here, the duality is realized in all fine print.

\subsection{Type--I blow--up model}

If we only give a VEV to $(\rep{1},\rep{1})_4$, the gauge group of the
blow--up model remains the same as on the heterotic orbifold. We identify
this case with blow--up model characterized by  
$V=(0^{4},1^{12})$ defined in ref.~\cite{Nibbelink:2007rd}. 
The spectra of the blow--up model and the type--I $\Intr_3$ orbifold
model~\cite{Angelantonj:1996uy,aldazabal_98} are identical, and given
in the second row of table~\ref{tb:Models}. Even though there is just a
single U(1), it does not factorize universally:
\equ{
(2\gp)^2 I_6^{blo} \!=\! \sm \frac {iF_1}{96} \Big(\!
 12 (iF_1)^2 \!+\! \tr(iF_{12})^2 \!-\! \tr(iF_8)^2 \!-\! \frac 18\tr R^2\!
\Big).
\non 
}
Because only the singlet takes a VEV, we make the following field
redefinitions of the twisted states 
\equ{
(\rep{1},\rep{1})_4 ~=~ e^T \, v~, 
\quad 
(\rep{8}_+, \rep{1})_{-2} ~=~ e^{-\frac 12 \, T} 
(\rep{8}_+,\rep{1})_{0}'~.  
}
The superpotential~\eqref{superP} gives the state
$(\rep{8}_+,\rep{1})_{0}'$ a regular mass. 
This transformation takes a twisted singlet to a localized axion
in the blow--up theory; from the type--I point of view this is the
twisted $RR$--axion. By computing the various anomaly polynomials we 
derive the identification of the axions 
\equ{
a^{non} ~=~ -\frac 1{16}\, a_T~, 
\quad 
a^{uni} ~=~ a^{het} ~+~ \frac 18\, a_T~. 
}
Hence the type--I $\Intr_3$ orbifold model coincides with a blow--up of
a heterotic  $\Intr_3$ orbifold.

\subsection{Blow--up model with two anomalous U(1)s}
\label{sc:twoU1model}

The second blow--up of the heterotic orbifold model is obtained by
giving VEVs to both $(\rep{8_+},\rep{1})_{\sm 2}$ and  
$(\rep{1},\rep{1})_4$. This induces further symmetry breaking of 
$\SO{8}$ to $\U{4}$, and therefore this model has two U(1)s. The
corresponding $\Cplx^3/\Intr_3$ blow--up model has a $\U{1}$ bundle 
characterized by $V= -\frac 12 (q+q')$ with the charge vectors 
$q = (1^{12}, 0^4)$ and $q' = (0^{12},3^4)$. The spectrum is given in
the last row of table~\ref{tb:Models}. The anomaly polynomial fails to
factorize:   
\equ{
(2\gp)^2 I^{blo}_6 =  
-\frac {2i F_1}3\Big( 
12(iF_1)^2 \!-\! 8 (iF_1')^2 
\!-\! 2 \tr(iF_4)^2 + 
\non \\[1ex]
\tr(iF_{12})^2 
\!-\! \frac 18 \tr\, R^2 
\Big)
+  2 iF_1' \Big( 4 (iF_1')^2 
\!+\! \tr(iF_4)^2 \!-\! \frac 18 \tr\, R^2
\Big), 
\non 
}
hence there are two anomalous U(1)s.  We can perform field
redefinitions to match the spectra of the orbifold and its blow--up version 
up to two singlets.  To this end we realize that the singlets 
$(\rep{1},\rep{1})_{\sm 2,\sm 2}$ and $(\rep{1},\rep{1})_{\sm 2,2}$
are obtained from $(\rep{8_+},\rep{1})_{\sm 2}$ after symmetry breaking. 
The redefined twisted states are: 
\equ{
(\rep{1},\rep{1})_{\sm 2, \sm 2} ~=~  e^T\, v\,,~~
(\rep{1},\rep{1})_{\sm 2, 2} ~=~ e^{-T}\, (\rep{1},\rep{1})_{\sm 4, 0}'\,. 
}
The singlets  $ (\rep{1},\rep{1})_{4, 0}$, the twisted orbifold
singlet, and 
$ (\rep{1},\rep{1})_{\sm4, 0}'$ missing in the blow--up pair up to
become massive. Also $(\rep{6},\rep{1})_{\sm 2,0}$ gets mass terms
because of Yukawa interactions involving $(\rep{1},\rep{1})_{4,0}$
that has a VEV as well.  The relation between the axions is given by:
\equ{
a^{non} ~=~ -\frac 1{16} \, a_T~, 
\qquad 
a^{uni} ~=~ a^{het} ~-~ \frac 1{16}\, a_T~. 
} 

\section{Acknowledgments}

We would like to thank A.\ Micu, F.\ Pl\"oger, M.\ Serone and P.\ Vaudrevange
for useful discussions. 
This work was partially supported by the European Union 6th framework
program MRTN-CT-2004-503069 "Quest for unification",
MRTN-CT-2004-005104 "Forces Universe", MRTN-CT-2006-035863
"Universe Net", SFB-Transregio 33 "The Dark Universe" and HE 3236/3-1
by Deutsche Forschungsgemeinschaft (DFG).

\bibliographystyle{paper}
{\small
\bibliography{paper}
}


\end{document}